%% file: phonon_rates_theory_v4.tex
\documentclass[floatfix,aps,a4paper,twocolumn,10pt,prb,showpacs]{revtex4-1}

\usepackage{graphicx}
\usepackage{amssymb}
\usepackage{amsmath}
\usepackage{psfrag}
\usepackage{epsfig}
\usepackage{hyperref}
\hypersetup{breaklinks}

\usepackage{xspace}

\newcommand{\ket}[1]{| #1 \rangle}

\newcommand{\rb}[1]{\left( #1 \right)}
\newcommand{\ew}[1]{\langle #1 \rangle}
\newcommand{\beq}{\begin{eqnarray}}
\newcommand{\eeq}{\end{eqnarray}}
\newcommand{\eq}[1]{Eq.~(\ref{#1})}
\newcommand{\fig}[1]{Fig.~\ref{#1}}
\newcommand{\secref}[1]{Sec.~\ref{#1}}
\newcommand{\bs}[1]{\boldsymbol{#1}}

\newcommand{\etal}{{\em et al.}\xspace}

\newcommand{\citer}[1]{{Ref.~\onlinecite{#1}}}

\begin{document}
\title{Phonon emission and arrival times of electrons from a single-electron source}
\author{C.~Emary, A.~Dyson}
\affiliation{
  Department of Physics and Mathematics,
  University of Hull,
  Kingston-upon-Hull,
  HU6 7RX,
  United Kingdom
}
\author{Sungguen~Ryu, H.-S.~Sim}
\affiliation{
  Department of Physics, Korea Advanced Institute of Science and Technology, Daejeon 305-701, Republic of Korea
}

\author{M.~Kataoka}
\affiliation{
 National Physical Laboratory, Hampton Road, Teddington, Middlesex TW11 0LW, United Kingdom
}

\date{\today}
\begin{abstract}
  In recent charge-pump experiments, single electrons are injected into quantum Hall edge channels at energies significantly above the Fermi level.
  We consider here the relaxation of these hot edge-channel electrons
  through longitudinal-optical phonon emission.
  Our results show that the probability for an electron in the outermost edge channel to emit one or more phonons en route to a detector some microns distant along the edge channel suffers a double-exponential suppression with increasing magnetic field.
  This explains recent experimental observations.
  We also describe how the shape of the arrival-time distribution of electrons at the detector reflects the velocities of the electronic states post phonon emission.  We show how this can give rise to pronounced oscillations in the arrival-time-distribution width as a function of magnetic field or electron energy.
\end{abstract}
\maketitle
%%%%%%%%%%%%%%%%%%%%%%%%%%%%%%%%%%%%%%%%%%%%%%%%%%%%%%%%%%%%%%%%%%%%%%%%%

\section{Introduction \label{SEC:Intro}}

Single-electron sources have recently been realised with a number of different technologies such as a driven mesoscopic capacitors \cite{Feve2007}, quantum-dot charge pumps \cite{Blumenthal2007,Kaestner2008,Kataoka2011,Giblin2012,Fletcher2012,Fletcher2013,Waldie2015}, and surface acoustic waves \cite{Kataoka2009,McNeil2011,Hermelin2011}.  Shaped voltage pulses have also been used to generate single Levitons \cite{Dubois2013}.
These sources enable electronic analogues of fundamental quantum-optics experiments \cite{Bocquillon2012,Ubbelohde2015,Freulon2015}, and hold great promise for future application, in particular as a current standard \cite{Pekola2013}.
For the full potential of these sources to be realised, however, we need an understanding of the relaxation  and decoherence processes that affect their single-electron outputs \cite{Degiovanni2009,Sueur2010, Ferraro2013,Ferraro2014}.

The focus of the current paper is the effect of longitudinal-optical (LO) phonon emission by the hot electrons originating from the charge pumps of Refs.~\onlinecite{Kataoka2011,Giblin2012,Fletcher2012,Fletcher2013,Waldie2015}.
In these systems, single electrons are emitted by a dynamically-driven quantum dot into quantum-Hall edge channels. The emission energy of these electrons can be controlled by adjusting the dot potentials \cite{Leicht2011,Fletcher2013} and can be set far above the Fermi sea.  By means of an adjustable detector barrier placed some $2$--$5\mu$m downstream of the emitter, these experiments offer both energy- and time-resolved detection of the electrons\cite{Fletcher2013,Waldie2015}.

In \citer{Fletcher2013} it was reported that, at certain magnetic field strengths, a significant fraction of the electrons arrive at the detector with an energy that is some integer-multiple of $\sim\!\!36$meV less than their energy at emission.  Since $36$meV corresponds approximately to the energy of LO phonons in GaAs \cite{Taubert2011}, it was concluded that these electrons had emitted one or more LO phonons en route to detector.  Moreover, Fletcher \etal  \cite{Fletcher2013} report that whilst these ``phonon replica'' features are pronounced at lower fields ($B=6$T), they are scarcely visible at high fields ($B=12$T).  A similar observation was made for different samples in \citer{Waldie2015}.

In this paper we offer an explanation of this transition based on the localisation properties of edge-channel wave functions in a magnetic field.  Using a Fr\"{o}hlich Hamiltonian \cite{Froehlich1954, Mahan2000}, we calculate scattering rates out of individual edge-channel states as a function of field and energy of emission.  From this we obtain the probability of electrons emitting $m=0,1,2,\ldots$ LO phonons before reaching the detector.  For electrons emitted into the outermost edge channel, we find an abrupt transition, essentially from zero to one, in the probability that no phonons are emitted.
We also discuss phonon emission by electrons in edge channels other than the outermost, and describe conditions under which anomalously large values of the relaxation rates can occur.  

We then go on to consider the distribution of electron arrival times at the detector. This distribution is similar in concept to the waiting time distribution, which has been studied for time-independent transport in Coulomb-blockade systems \cite{Brandes2008} and also for a dynamic single-electron emitter \cite{Albert2011}.  
Here, we calculate the arrival-time distributions (ATDs) for electrons having emitted different numbers of phonons and show how energy loss leads to an increase in the widths of the phonon-replica distributions.  We also predict that at lower fields the widths of these distributions show pronounced oscillations as a function of emission energy or magnetic field.
This effect originates from the scattering of electrons into different edge channels as the field or energy is changed.

This paper is structured as follows. In \secref{SEC:QHEmodel} we recap the properties of quantum-Hall edge states with parabolic transverse confinement and in 
\secref{SEC:rates} we derive an expression for the scattering rate between these states.  The ATDs are discussed in \secref{SEC:ATD}, before we finish with discussions in \secref{SEC:discussions}.

%%%%%%%%%%%%%%%%%%%%%%%%%%%%%%%%%%%%%%%%%%%%%%%%%%%  %%%%%%%%%%%%%%%%%%%%%%
\section{Electron states \label{SEC:QHEmodel}}

The charge pumps in question emit single electrons into the edge channels of a two-dimensional electron gas in the quantum-Hall regime.
We model the behaviour of these electrons in $xy$-plane with the effective-mass Schr\"odinger equation \cite{Datta1997} ($e>0$)
\beq
  H
%   (\mathbf{r}) 
= 
  %E_c + 
  \frac{1}{2m_e^*}
  \rb{
%   \left[
    i \hbar \bs{\nabla} -e \mathbf{A}
%   (\mathbf{r})
%   \right]
   }^2
  + U(y) 
%   + U(z)
  \label{EQ:h}
  ,
\eeq
with $m_e^*$ the effective electron mass,  $\mathbf{A}$ the vector potential, and $U(y)$ the confinement potential transverse to the transport direction.
With magnetic field $B$ perpendicular to the plane, the vector potential in the Landau gauge reads  $\mathbf{A} = - B y \,\hat{\mathbf{i}}$, with $\hat{\mathbf{i}}$ a unit vector in the $x$-direction.
We consider a parabolic confinement with confinement energy $\hbar  \omega_y$ such that
$
  U(y) = \frac{1}{2} m_e^* \omega_y^2 y^2
$.
% \label{EQ:Uy}
We discuss the limits of this model in describing the experiments of Refs.~\onlinecite{Fletcher2013,Waldie2015} at the end of this section.

The eigenfunctions of $H$ are plane waves, with wave number $k$, in the $x$ direction, and harmonic-oscillator eigenfunctions with quantum number $n=0,1,2,\ldots $ in the $y$ direction (see Appendix \ref{SEC:APQHstates}).  The transverse wave functions are localised about a guiding-centre coordinate
\beq
  y_\text{G}(k) =  \frac{\omega_c^2}{\Omega^2}\frac{\hbar {k} }{eB} 
  ,
\eeq
and have a characteristic width
\beq
  l_\Omega = \sqrt{\frac{\hbar}{m_e^* \Omega}}
  .
\eeq
Here, $\Omega$ is the composite frequency
$
  \Omega = \sqrt{\omega_y^2 + \omega_c^2}
$
with $
  \omega_c = \frac{|eB|}{m_e^*}
$,
the cyclotron frequency.
The energies of the eigenstates read
\beq
  E_{nk} &=& 
  \hbar\Omega\left\{
    n + \frac{1}{2}
    +  \frac{1}{2}
      \left[
        \frac{\omega_y y_\text{G}(k)}{ \omega_c l_\Omega}
      \right]^2
  \right\}
  \label{EQ:disp1}
  ,
\eeq
with corresponding velocities
\beq
  v_{nk} =
  \frac{1}{\hbar} \frac{\partial E_{nk}}{\partial k}
  =
  \frac{\omega_y^2}{\omega_c} y_\text{G}
  \label{EQ:vel}
  .
\eeq

Recent measurements \cite{Kataoka2015} of the velocities of the electrons emitted by the charge pumps show that, close to edge of the sample, the transverse potential is well approximated by the quadratic form employed here. Across the interior of the sample, however, the potential is expected to be essentially flat. The experimental potential is therefore an open parabola, rather that the closed one we consider here.
Nevertheless, we expect the eigenfunctions of the closed parabola to provide a good approximation to those of the open one, if their displacement from the origin is significantly greater than their spatial extent, i.e. when $y_\text{G} \gg l_\Omega$.  This implies that the energy of the electron above its subband bottom should satisfy
\beq
  E-\hbar\Omega\rb{n+\frac{1}{2}}
  \gg
  \frac{\hbar \Omega}{2} \rb{\frac{\omega_y}{\omega_c}}^2
  \label{EQ:Emodelconst}
  .
\eeq
This holds true for most of the results reported here. In the cases where it does not, we will argue that our results still give a qualitative guide to experiment.

%%%%%%%%%%%%%%%%%%%%%%%%%%%%%%%%%%%%%%%%%%%
\section{Phonon relaxation rates \label{SEC:rates}}

The scattering of quantum-confined electrons by phonons, both with \cite{DasSarma1980,Larsen1984,Telang1993,Taubert2011} and without \cite{Riddoch1984,Bockelmann1990} magnetic field, has been studied extensively.  These previous studies, however, have focused on macroscopic properties such as conductance or optical absorption.  In contrast, our analysis here concerns the fate of single electrons and scattering rates between individual edge-channel states.

We describe the interaction between electrons and LO phonons with the
Fr\"{o}hlich Hamiltonian \cite{Froehlich1954, Mahan2000} which, in terms of the electronic states described above, can be written as
\beq
  V_\text{ep} &=&
  \sum_{n n'}  \sum_{k k'}
  \sum_{\mathbf{q}}
    \Lambda_{n'n}^{k'k}(\mathbf{q})
    c_{n'k'}^\dag c_{nk}
    \rb{a_{-\mathbf{q}}^\dag  + a_\mathbf{q}}
  \label{EQ:V}
    .
\eeq
Here $c_{nk}$ is the annihilation operator for electrons with quantum numbers $n$ and $k$, $ a_\mathbf{q}$ is the annihilation operator for bulk LO phonons with (three-dimensional) wave vector $\mathbf{q}$, and $\Lambda_{n'n}^{k'k}(\mathbf{q})$ is the appropriate matrix element, proportional to the dimensionless Fr\"ohlich coupling constant $\alpha$.
The form of this matrix element along with some technical details on the following calculation are discussed in  Appendix~\ref{SEC:APep}.   
We assume that the phonons are dispersionless and have energy $\hbar \omega_\text{LO}$.

We consider the zero-temperature limit and phonon emission only. 
Since in GaAs the coupling constant is small ($\alpha \approx 0.068$), we work to lowest order in $\alpha$ and calculate the effects of coupling to the phonons via Fermi's golden rule \cite{Ridley2009}. 
This gives the rate of transition from the state in subband $n$ with energy $E$ to a state in subband $n'$ with energy $E-\hbar \omega_\text{LO}$ to be
\beq
  \Gamma_{n'n}(E)
  &=&  
 \frac{\alpha \Omega \omega_\text{LO}}{2 \pi\omega_y}
  \,
  \sqrt{\frac{\hbar \omega_\text{LO}}{\Delta_{n'}}}
  \Theta\rb{\Delta_{n'}}
  I_{n'n}\rb{\delta_G}
  \label{EQ:Ifull}
  . 
\eeq
Here,
$
   \Delta_{n'}(E)
   \equiv E - \hbar \Omega \rb{n'+\textstyle{\frac{1}{2}}} - \hbar\omega_\text{LO}
$
is the energy taken up by the motion along the edge channel after the transition, $\Theta$ is the unit-step function, and
$I_{n'n}$ is a one-dimensional integral, the argument of which is the relative change in guide centre,
\beq
  \delta_G = \frac{y_\text{G} - y_\text{G}'}{l_\Omega},
\eeq
with $y_\text{G}$ associated with the initial state (subband $n$) and $y_\text{G}'$ with the final state (subband $n'$).
The full form of this one-dimensional integral is discussed in Appendix~\ref{SEC:APep}.  However, when the initial energy of the electron is the dominant energy scale, we obtain the approximate form
\beq
  I_{n'0}(\delta_G)
  &  \approx&
  \pi^{3/2}
  \frac{\omega_c}{\Omega}
  \frac{1}{n'!} 
  \rb{\frac{\delta_G^2}{2}}^{n'-\frac{1}{2}}
  e^{-\frac{1}{2}\delta_G^2}
  \label{EQ:Iapp}
  ,
\eeq
where, for simplicity of presentation, we quote the result for $n=0$ only. 
Thus, the rates are dominated by an exponential dependence on the distance between guiding centres before and after scattering.  This dependence arises from the overlap of the transverse wave functions.  
In emitting a phonon, an electron loses energy and, if starting in the outermost subband, its $k$-value is correspondingly reduced.  Since the guiding centre of the edge-channel wave functions is proportional to $k$, states before and after emission are then necessarily separated in the $y$ direction. At large field and/or energies, the overlap of these two wave functions is through their exponential tails, and hence the form of \eq{EQ:Iapp}.

Considering first the transition within the outermost edge-channel ($n=n'=0$), we have
\beq
  \delta_G^2
  =
  \frac{2\omega_c^2}{\omega_y^2\hbar \Omega}
  \left[
    \sqrt{E-\frac{1}{2} \hbar \Omega}
    -
    \sqrt{E - \hbar \omega_\text{LO}-\frac{1}{2} \hbar \Omega}
  \right]^2
  \nonumber
  .
\eeq
If the energies of the problem are ordered $E-\hbar \omega_\text{LO} \gg \hbar \omega_c \gg \hbar \omega_y$, this simplifies such that we may write
\beq
  \Gamma_{00}(E)
  \sim
  \exp\left[
    - B/B_\text{T}
  \right]
  \label{EQ:Gamma_simple}
  ,
\eeq
with 
\beq
  B_\text{T} = 
  \frac{m \hbar \omega_y^2 }{e E} \rb{1-\sqrt{1-\frac{\hbar \omega_\text{LO}}{E}}}^{-2}
  \label{EQ:Gamma_BT}
  .
\eeq
This transition field increases with increasing electron energy and also with increasing confinement. For higher field strengths, terms proportional to $B^2$ start to play a role in the exponent.

Considering the case for general $n$ and $n'$, in the same approximation as above and with $E \gg \hbar \omega_\text{LO}$, the exponential part of the rate reads
\beq
  \Gamma_{n'n}(E)
  &\sim&
  \exp\bigg[
    - \frac{1}{4E}
   \frac{\rb{\hbar \omega_c}^2}{\hbar\Omega}
   \rb{\frac{\omega_\text{LO}}{\omega_y}}^2
   \nonumber\\
   &&
   ~~~ ~~~~\times
%     \frac{1} {2E}
    \left\{
      1 + \frac{\Omega}{\omega_\text{LO}}\rb{n'-n}
    \right\}^2
  \bigg]
  .
\eeq
The dependence on $n'-n$ here means that for large fields, the  inwards off-diagonal transitions, $n'>n$, are increasingly more suppressed than the diagonal ones. Thus starting in the outermost channel, the $n=0$ to $n'=0$ transition will dominate at high field.  For $\Omega \lesssim \omega_\text{LO}$, however, rates other than the diagonal ones will contribute.

%%%%%%%%%%%%%%%%%%%%%%%%%%%%%%%%%%%%%%%%%%%%%%%%%%%%%%%%%%%%%%%%
\begin{figure}[t]
  \psfrag{Gamma1}{$\Gamma_0\,\left[\text{ps}^{-1}\right]$}
  \psfrag{Gamma2}{$\Gamma_0\,\left[\text{ps}^{-1}\right]$}
  \psfrag{B}{$B~\text{[T]}$}
  \psfrag{E=50meV}{$50$meV}
  \psfrag{E=100meV}{$100$meV}
  \psfrag{E=150meV}{$150$meV}
  \begin{center}
     \includegraphics[width=\columnwidth,clip=true]{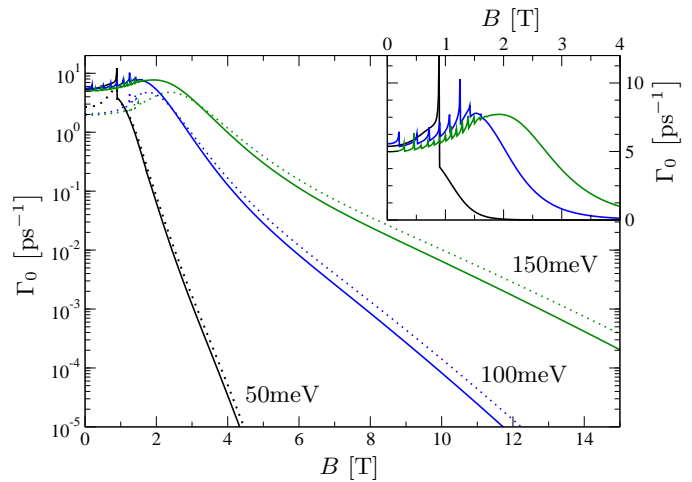}
  \end{center}
  \caption{
    The total phonon-induced scattering rate $\Gamma_0 =\Gamma_0(E)$ out of the $n=0$ edge channel as a function of magnetic field for several initial electron energies, $E=50,100,150$meV.
    The main panel shows results out to $B=15$T for both the full expression of \eq{EQ:Ifull} (solid lines) as well as the approximate form of \eq{EQ:Iapp} (dashed lines).
    The inset shows the low field region.
    Parameters for the calculation were:   
    transverse confinement energy of $\hbar \omega_y = 2.7$meV;
    electron-phonon coupling constant: $\alpha=0.068$;
    effective mass: $m_e^* =  0.067 m_e$; 
    and
    phonon energy: $\hbar \omega_\text{LO} = 36$meV.  We also assumed a confinement distance of $a=5$nm perpendicular to the plane (see Appendix~\ref{SEC:APQHstates}).
    \label{FIG:rates_m_eq_0}
  }
\end{figure}
%%%%%%%%%%%%%%%%%%%%%%%%%%%%%%%%%%%%%%%%%%%%%%%%%%%%%%%%%%%%%%%%
 
The total scattering rate out of state $n$ is simply the sum $\Gamma_n(E) = \sum_{n'} \Gamma_{n'n}(E)$.  Results for $n=0$, using both the exact integral $I_{n'0}$ and the approximate form of \eq{EQ:Iapp}, are shown in \fig{FIG:rates_m_eq_0}.
This figure shows that the rate exhibits an approximately exponential drop across most of the experimentally-accessible magnetic-field range and that \eq{EQ:Iapp} provides a decent account of this behaviour.
The inset of \fig{FIG:rates_m_eq_0} shows the total rate at low fields. 
We see that the decay rate has a maximum value in the range 5\,--10ps$^{-1}$ and occurs for $B>0$.  The rate at low fields also exhibits a series of peaks as a function $B$ that arise from the density of states factor, $\Delta_{n'}^{-1/2}$, in \eq{EQ:Ifull}.
At these points, \eq{EQ:Emodelconst} does not hold and the closed parabola is no longer an accurate model of the experimental potential. 
Missing from the current description are the bulk states that occur in the flat region of the potential.  Due to the magnetic field, however, these additional states will be dispersionless and the spectrum of the open-parabola system will still consist of a set of distinct subbands. Thus, even though the positions and strengths of these peaks will be modified in the open potential, the essential ingredient behind this behaviour remains.  We thus expect the behaviour described here to be qualitatively similar to that observable in experiment.

\begin{figure}[t]
  \begin{center}  
  \psfrag{Gamma}{$\Gamma_n\,\left[\text{ps}^{-1}\right]$}
  \psfrag{K}{$k~\left[\text{nm}^{-1}\right]$}
  \psfrag{B}{$B~\text{[T]}$}
  \psfrag{m=0}{$n=0$}
  \psfrag{m=1}{$n=1$}
  \psfrag{m=2}{$n=2$}
  \psfrag{m = 0}{$\!\!n=0$}
  \psfrag{m = 1}{$\!\!n=1$}
  \psfrag{m = 2}{$\!\!n=2$}
     \includegraphics[width=\columnwidth,clip=true]{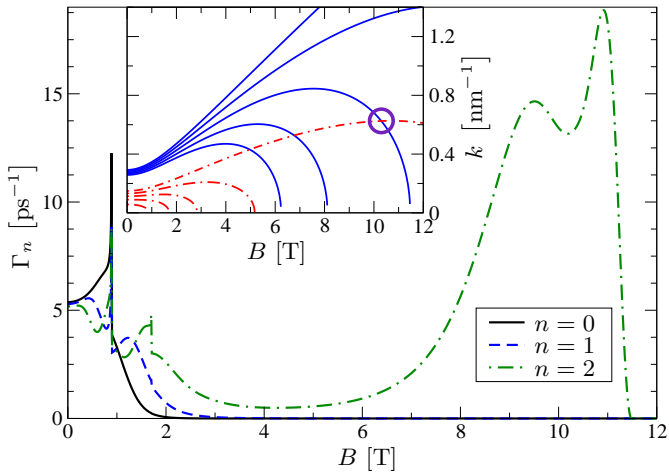}
  \end{center}
  \caption{
    {\bf Main panel:} Total scattering rate $\Gamma_n =\Gamma_n(E)$ out of the $n=0,1,2$ edge channel as a function of magnetic field with an initial electron energy of $E=50$meV. 
    The most striking feature is the anomalously high rate for the $n=2$ state centered around a field of $B\approx10$T.
    {\bf Inset:}
    This phenomenon can be understood by considering the wave numbers involved in scattering as a function of magnetic field.  The solid blue lines show the wave numbers of initial states at $E=50$meV as a function of magnetic field.  The uppermost line is from the $n=0$ subband, with lines progressing downwards corresponding to $n=1,\ldots,4$.  The dashed red lines show the same thing but at an energy of $E-\hbar\omega_\text{LO}=14$\,meV.  A crossing between the $n=2$ initial-state line and the $n=0$ final-state line occurs around $B=10.3$T (circled).  At this point the momentum of the electron in the transport direction is conserved by the scattering and this gives rise to the enhanced rate observed in the main panel.
    Parameters the same as \fig{FIG:rates_m_eq_0}.
    \label{FIG:rates_m_ge_0}
  }
\end{figure}

\fig{FIG:rates_m_ge_0} shows the total scattering rate out of states other than the $n=0$ level. Generically, these rates show a behaviour similar to the $n=0$ case.  In certain circumstances, however, these rates can attain anomalously high values, see  e.g. the $n=2$ result in \fig{FIG:rates_m_ge_0}a around $B= 10$T.
These features occur whenever initial-state and final-state lines cross in a plot of wave number versus field (see the inset of \fig{FIG:rates_m_ge_0}b).  At these points, the momentum of the electron in the $x$ direction is conserved.  From the $1/q^2$-structure of the integrals, this gives rise to enhancement in the scattering rate.

%%%%%%%%%%%%%%%%%%%%%%%%%%%%%%%%%%%%%%%%%%%%%%%%%%%%%%%%%%%%%%%%
%%%%%%%%%%%%%%%%%%%%%%%%%%%%%%%%%%%%%%%%%%%%%%%%%%%%%%%%%%%%%%%%

%%%%%%%%%%%%%%%%%%%%%%%%%%%%%%%%%%%%%%%%%%%%%%%%%%%%%%%%%%%%%%%%%%%%%%%%%%%%%%%%%%%%%%%
%%%%%%%%%%%%%%%%%%%%%%%%%%%%%%%%%%%%%%%%%%%%%%%%%%%%%%%%%%%%%%%%%%%%%%%%%%%%%%%%%%%%%%%
%%%%%%%%%%%%%%%%%%%%%%%%%%%%%%%%%%%%%%%%%%%%%%%%%%%%%%%%%%%%%%%%%%%%%%%%%%%%%%%%%%%%%%%
\section{Arrival-time distribution\label{SEC:ATD}}

We now consider the time of arrival of electrons at a detector situated at a position $x_\text{D}$ from the emitter along the edge channel.  
Let $E^{(0)}$ be the energy of the electrons at emission and let us denote as $A^{(m)}(x_\text{D},\tau)$ the distribution of arrival times of electrons with energy $E^{(m)} = E^{(0)}-m\hbar \omega_\text{LO}$, i.e. ones having emitted exactly $m$ phonons.  The normalisation of $A^{(m)}$ is such that $P^{(m)} (x_\text{D}) \equiv \int_0^\infty d\tau A^{(m)}(x_\text{D},\tau)$ is the total probability of $m$ phonons being emitted {\em en route} and $\sum_{m=0} P^{(m)} (x_\text{D}) =1$.
For each value of $m$, we define the mean arrival time
\beq
  \ew{\tau^{(m)}}
  \equiv
  \frac{  
    \int_0^\infty d \tau
    \,
    \tau A^{(m)}(x_\text{D},\tau)
  }
  {
    P^{(m)}(x_\text{D})
  }
  ,
\eeq
and width 
\beq
  \Delta \tau^{(m)}
  \equiv
  \sqrt{
    \frac{
      \int_0^\infty d \tau
      \,
      \left[
	\tau^2 - \ew{\tau^{(m)}}^2
      \right]
      A^{(m)}(x_\text{D},\tau)  
    }{
      P^{(m)}(x_\text{D})
    }
  }
  .
\eeq

To calculate the ATDs, we should consider that the electron is emitted as a wave packet with a range of $k$ values. In principle, this affects the time evolution of the electron not only through the dispersion of \eq{EQ:disp1}, but also through the energy (and hence wavenumber) dependence of the scattering rate.  A quantum-mechanical treatment of the ATD that addresses these issues is discussed in Appendix \ref{SEC:APQM}.  
In this treatment, we assume that the initial spatial distribution of the electron is a Gaussian wave packet with a spatial extent $\gtrsim 1\mu$m, which is a reasonable assumption for the situation described in \citer{Fletcher2013,Waldie2015}.   In this case, we find that the relative spread in wave number $k$ is small enough that the dispersion of the wave packet can be neglected over relevant source-detector distances.  The variation of the phonon rates, the guide-centre positions and the velocities over the relevant $k$ range are also negligibly small.
In the following, then, we neglect dispersion and employ a semi-classical description of the dynamics.  We note that where the initial extent of the wave packet $\lesssim100$nm,  we would expect $k$-dependent effects to become significant.

\subsection{Semi-classical dynamics}

Let us denote as $\varrho_{n}^{(m)}(x,t)$ the classical probability distribution to find at time $t$ an electron at position $x$ of edge channel $n$  given that it has emitted $m$ phonons.  In a semi-classical picture, we consider an electron in edge channel $n$ to have a well-defined velocity, irrespective of its spatial distribution.  We will label quantities with the phonon-number $m$ as a proxy for the energy and thus write $v_n^{(m)}$ for the velocity of the electron in the $n$th subband with energy $E^{(m)} = E^{(0)}-m \hbar \omega_\text{LO}$.  We label the rates with the $m$-value of the starting state, rather than the energy: $\Gamma_{n'n}^{(m)} = \Gamma_{n'n}\rb{E^{(m)}}$.

For the probabilities, we write down a set of coupled Boltzmann-like equations with drift and scattering terms
\beq
  \frac{\partial}{\partial t} \varrho_{n}^{(m)}
%   (x,t)
  +
  v^{(m)}_{n} \frac{\partial}{\partial x} \varrho_{n}^{(m)}
%   (x,t)
  &=&
   -\rb{1-\delta_{m,M}}
  \Gamma_{n}^{(m)}  
  \varrho^{(m)}_{n}
  \nonumber\\
  &&
  +
  \rb{1-\delta_{m,0}}
  \sum_{n'} \Gamma_{nn'}^{(m-1)} 
  \varrho^{(m-1)}_{n'}
  .
  \nonumber\\
  \label{EQ:EOM_Pall}
\eeq
Here $M$ is the maximum number of phonons that an electron can emit before its energy fails to less than $\hbar \omega_\text{LO}$ above the bottom of the $n=0$ subband, from which point no further emission is possible.
In terms of these probabilities, the ATD at energy $E^{(m)}$ is simply related to the current as \cite{Muga2000}
\beq
  A^{(m)}(x_\text{D}, \tau) 
  &=& 
  \sum_n A_n^{(m)}(x_\text{D}, \tau)
  \nonumber\\
  & =&
  \sum_n v_n^{(m)} \varrho_{n}^{(m)}(x_\text{D},t=\tau)
  .
\eeq
In the first line here we have written $A^{(m)}$ as a sum over the contributions from the individual edge channels, $A^{(m)}_n$.

\subsection{The survival probability \label{SEC:survprob}}

\begin{figure}[t]
  \psfrag{P0}{$P^{(0)}$}
  \psfrag{B}{$B~\text{[T]}$}
  \psfrag{Bhalf}{$B_{1/2}~\text{[T]}$}
  \psfrag{tbydt}{$\Delta \tau^{(0)} / \Delta \tau_\text{i}$}
  \psfrag{E}{$E$~[meV]}
  \psfrag{E=50meV}{\!\!$50$meV}
  \psfrag{E=100meV}{\!\!$100$meV}
  \psfrag{E=150meV}{\!\!$150$meV}
  \psfrag{x=5um}{\!$5\mu$m}
  \psfrag{x=10um}{\!$10\mu$m}
  \psfrag{x=15um}{\!$15\mu$m}
  \psfrag{(a)}{\textbf{(a)}}
  \psfrag{(b)}{\textbf{(b)}}
  \psfrag{(c)}{\textbf{(c)}}
  \begin{center}
     \includegraphics[width=\columnwidth,clip=true]{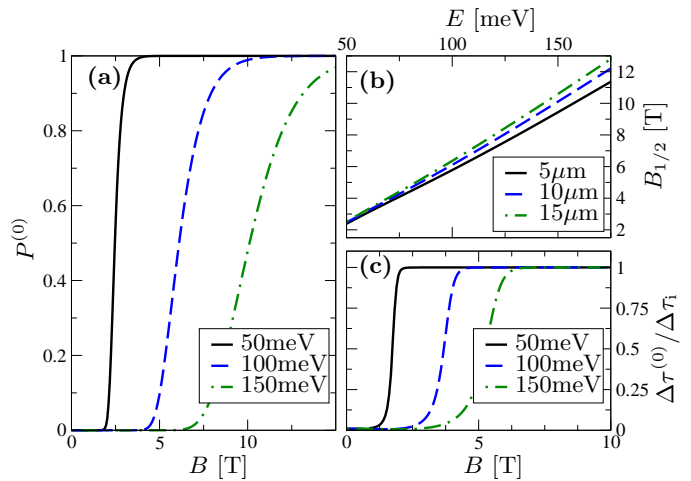}
  \end{center}
  \caption{
    \textbf{(a)} The survival probability, $P^{(0)}$, for electrons to reach a detector at $x_\text{D} =10\mu$m without emitting a phonon for initial energies $E^{(0)} = 50,100,150$meV.
    A rapid transition from zero to unity is observed in the experimentally-relevant range of magnetic fields.
    \textbf{(b)} The field value $B_{1/2}$ at which the survival probability reaches one half as a function of initial energy $E$. Results are shown for $x_\text{D} = 5,10,15\mu$m. 
    \textbf{(c)}  The width of the $m=0$ arrival-time distribution, $\Delta \tau^{(0)}$, in units of the initial width $\Delta \tau_\text{i}$ (Same energies and detector position as part (a)).  Above $B_{1/2}$, the width plateaus to its value at emission.  
    Parameters the same as \fig{FIG:rates_m_eq_0}.
    \label{FIG:P0}
  }
\end{figure}

The first quantity in which we are interested is the survival probability, $P^{(0)}(x_\text{D})$, which is the probability that the electron reaches the detector without emitting any phonons.  We assume that the electron is emitted into the outermost edge channel with a starting probability density 
$
  \varrho_{0}^{(0)}(x,0)  = f(x)
$.
Solution of \eq{EQ:EOM_Pall} with $m=n=0$ gives simply
\beq
 \varrho_{0}^{(0)}(x,t) = 
  e^{- \Gamma_0^{(0)} t}
  f(x - v_0^{(0)} t)
  \label{EQ:P0damped}
  ,
\eeq
which represents a wave packet travelling with velocity $v_0^{(0)}$, damped at a rate $\Gamma_0^{(0)} $.
We assume that the initial distribution is a Gaussian of spatial width $\sigma$ and central coordinate $x_0$.
In \citer{Waldie2015}, it was determined that the electron wave packet was emitted over some fixed time interval $\Delta \tau_\text{i}$.  This gives the initial width of the wave packet to be $\sigma = v^{(0)}_0 \Delta \tau_\text{i}$.  Since the wave packet begins to form at $t=0$, we set the central coordinate $x_0 = -2\sigma$ such that $\varrho_0(x \ge 0,0)\approx 0 $ for $x\ge 0$. \citer{Waldie2015} describes $30$ps as an overestimate of width, and we shall take 20ps in our numerical calculations.
By integrating \eq{EQ:P0damped} with this Gaussian Ansatz, we find a survival probability
\beq
   P^{(0)}(x_\text{D}) &=&
  \frac{1}{2} 
  \exp\left\{
    -\frac{\Gamma_0^{(0)} (x_\text{D}-x_0)}{v_0^{(0)}}+ 
    \frac{1}{2}\rb{\frac{\sigma \Gamma_0^{(0)}}{v_0^{(0)}}}^2
  \right\} 
  \nonumber\\
  &&
  \times
  \text{erfc}
  \left\{
    \frac{\sigma^2 \Gamma_0^{(0)}-v_0^{(0)} (x_\text{D}-x_0)}{\sqrt{2} \sigma v_0^{(0)}}
  \right\}
  .\label{EQ:P00full}
\eeq
This result is shown in \fig{FIG:P0}a with rates and velocities calculated as in the preceding sections. We see that the survival probability undergoes a rapid transition from close to zero for low fields to close to unity for high field.  
In the limit $(x_\text{D}-x_0)/\sigma \gg 1, \sigma \Gamma_0^{(0)}/v_0^{(0)}$, the degree of scattering experienced by the wave packet as it passes the point $x_\text{D}$ is negligible,
and we find
\beq
  P^{(0)} (x_\text{D})\approx 
  \exp \left\{- \frac{\Gamma_0^{(0)} (x_D-x_0)}{v_0^{(0)}}\right\}
  \label{EQ:P00approx}
  ,
\eeq
which represents a simple exponential decay in time evaluated at the mean-time of arrival at the detector.  
Since, for large $B$ at least, the rate $\Gamma^{(0)}_{0}$ behaves as in \eq{EQ:Gamma_simple}, the result of \eq{EQ:P00approx} is a {\em double-exponential} suppression on the survival probability as the magnetic field decreases.
From \eq{EQ:Gamma_simple} we also see that the magnetic field at which the survival probability reaches one-half, $B_{1/2}$, is roughly proportional to $B_\text{T}$.  Complete results for this cross-over field are shown in \fig{FIG:P0}b as a function of energy.
We note that at low field, the rate $\Gamma_0^{(0)}$ is high enough that the wave packet is almost entirely suppressed before it can reach the detector. 
In this case, \eq{EQ:P00approx} ceases to be a good approximation.  Rather, in the limit $\sigma \Gamma_0^{(0)}/v_0^{(0)} \gg 1, (x_\text{D}-x_0)/\sigma$, 
we obtain
\beq
   P^{(0)} (x_\text{D})\approx 
   \frac{v_0^{(0)}}{\Gamma_0^{(0)}}f(x_\text{D})
   \label{P00:bigG}
   ,
\eeq
such that only the exponentially-small tail of the distribution contributes in this limit.

%%%%%%%%%%%%%%%%%%%%%%%%%%%%%%%%%%%%%%%%%%%%%%%%%%%%%%%%%%%%%%%%
%%%%%%%%%%%%%%%%%%%%%%%%%%%%%%%%%%%%%%%%%%%%%%%%%%%%%%%%%%%%%%%%

\subsection{Distribution widths}

%%%%%%%%%%%%%%%%%%%%%%%%%%%%%%%%%%%%%%%%%%%%%%%%%%%%%%%%%%%%%%%%
\begin{figure*}[t]
  \psfrag{tps}{$t$\,[ps]}
  \psfrag{cack}{$E^{(0)}$\,[meV]}
  \psfrag{0}{$0$}
  \psfrag{100}{$100$}
  \psfrag{200}{$200$}
  \psfrag{40}{$40$}
  \psfrag{80}{$80$}
  \psfrag{120}{$120$}
  \psfrag{160}{$160$}
  \psfrag{m=0}{$m=0$}
  \psfrag{m=1}{$m=1$}
  \psfrag{m=2}{$m=2$}
  \begin{center}
    \includegraphics[width=\textwidth,clip=true]{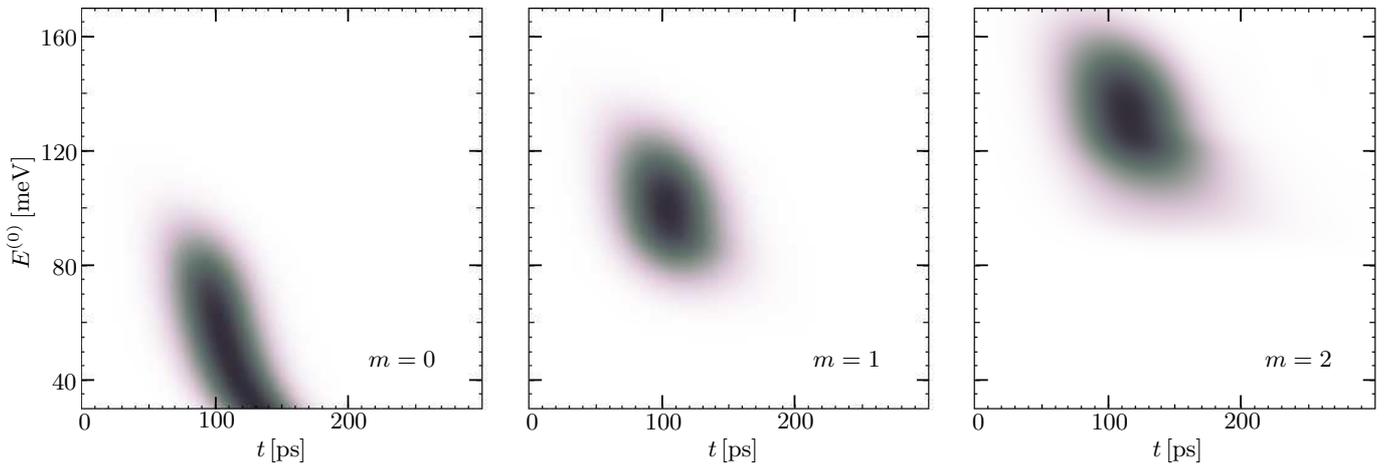}
  \end{center}
  \caption{
    Arrival-time distributions as function of time $t$ and initial electron energy, $E^{(0)}$.  The three panels show the distributions detected at energies $E^{(m)}$ with  $m=0,1,2$, i.e. with the electron having emitted $m=0,1,2$ phonons. 
    Here $B=5$T, $x_\text{D} = 10\mu$m, and the initial width was $\Delta \tau_\text{i}  = 20$ps.
    As the initial energy increases, the distribution moves from the $m=0$ to $m=1$ and then $m=2$ as one- then and two-phonon emission processes become possible.
    \label{FIG:AxtB5}
  }
\end{figure*}
%%%%%%%%%%%%%%%%%%%%%%%%%%%%%%%%%%%%%%%%%%%%%%%%%%%%%%%%%%%%%%%%

We now consider the complete ATDs and characterise them in terms of their  widths.  Considering first the $m=0$ case, the same approximations that lead to \eq{EQ:P00approx}, give the width of the $m=0$ ATD to be
\beq
  \Delta \tau^{(0)} = \frac{\sigma}{v_0^{(0)}}
  ,
\eeq
which is simply the typical time it takes for a wave packet of width $\sigma$ to move across the detector.  In the emission model such that $\sigma = v_0^{(0)} \Delta \tau_\text{i}$, we  obtain $\Delta \tau^{(0)} = \Delta \tau_\text{i}$ and the width of this distribution remains constant irrespective of decay process. This behaviour is observed at high fields in \fig{FIG:P0}c, but as $B$ decreases through the transition point, $B_{1/2}$, the width of the arrival-time distribution drops.  This is consistent with the picture of \eq{P00:bigG} that only the small fraction of the probability distribution close to the measurement point contributes in the strongly-damped regime.
We note that the fixity of the temporal width relies on the assumption that the electron is emitted over a constant time-window irrespective of other conditions.  If, for example, it were the initial spatial width $\sigma$, rather than the temporal width, that was fixed,  then $ \Delta \tau^{(0)}$ would show an approximately linear increase with field.

\begin{figure*}[t]
  \psfrag{tps}{$t$~[ps]}
  \psfrag{cack}{$E^{(0)}$\,[meV]}
  \psfrag{0}{$0$}
  \psfrag{200}{$200$}
  \psfrag{400}{$400$}
  \psfrag{40}{$40$}
  \psfrag{60}{$60$}
  \psfrag{80}{$80$}
  \psfrag{100}{$100$}
  \psfrag{m=0}{$m=0$}
  \psfrag{m=1}{$m=1$}
  \psfrag{m=2}{$m=2$}
  \begin{center}
  \includegraphics[width=\textwidth,clip=true]{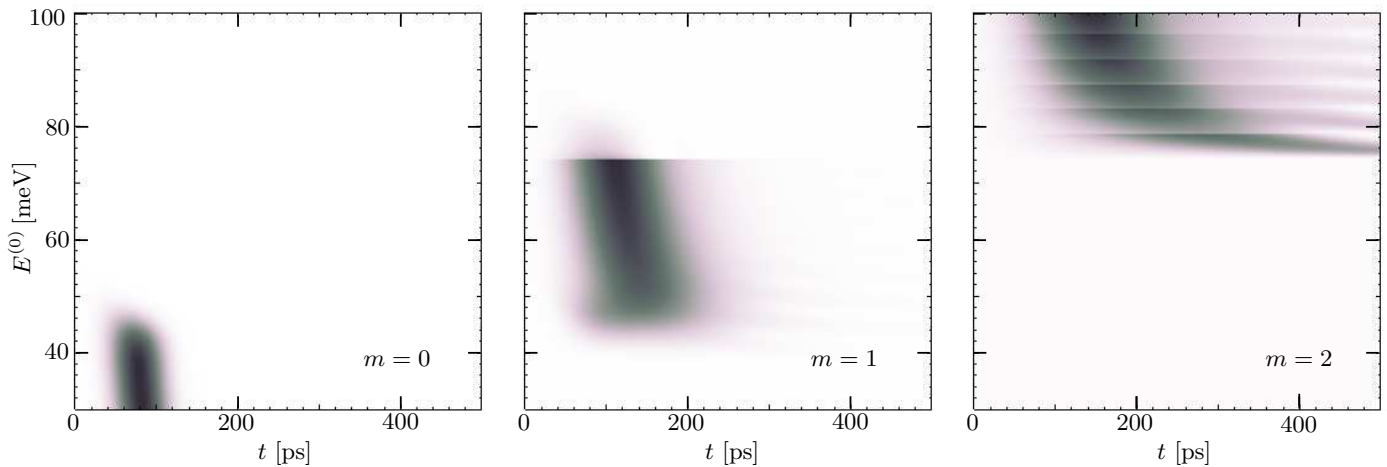}
  \end{center}
  \caption{  
    Same as \fig{FIG:AxtB5} but here with $B=2$T. The transitions here between the distributions with different $m$ are more abrupt. Furthermore, the distributions develop oscillations in their widths as a function of initial energy.
    \label{FIG:AxtB2}
  }
\end{figure*}

We next consider the distributions for $m>0$.  Results from the numerical solution of \eq{EQ:EOM_Pall} are shown in \fig{FIG:AxtB5} for $B=5$T and in \fig{FIG:AxtB2} for $B=2$T.  We start by discussing \fig{FIG:AxtB5}, as these results are indicative of what happens at higher field.  
Starting at low energy, the ATD for $m=0$ shows a strong peak given by \eq{EQ:P0damped}.  As the initial energy of the electron increases, this peak moves to shorter times as the velocity increases in line with \eq{EQ:vel}. 
At around 
$E^{(0)} = \frac{1}{2}\hbar\Omega + \hbar \omega_\text{LO}
\approx80
$meV, scattering out of this state becomes significant and population transfers to the $m=1$ state.  This continues until about 
$
E^{(0)}
= \frac{1}{2}\hbar\Omega + 2\hbar \omega_\text{LO}
\approx115
$meV, at which point the emission of two phonons becomes likely and the $m=2$ distribution develops.  The temporal widths of the distributions increase with increasing $m$, but narrow throughout their individual ranges as $E^{(0)}$ increases.  The distributions here are largely featureless.

%%%%%%%%%%%%%%%%%%%%%%%%%%%%%%%%%%%%%%%%%%%%%%%%%%%%%%%%%%%%%%%%
%%%%%%%%%%%%%%%%%%%%%%%%%%%%%%%%%%%%%%%%%%%%%%%%%%%%%%%%%%%%%%%%

A similar story unfolds at lower fields, \fig{FIG:AxtB2}, but the increase in width with $m$ here is more marked.  Moreover, a distinct structure evolves in the $m>0$ distributions.
Whilst the transition from $m=0$ to $m=1$ distribution is gradual, the $m=1$ distribution undergoes an abrupt cut-off around 
$
  E
  = \frac{1}{2}\hbar \Omega + 2 \hbar \omega_\text{LO}
  \approx74
$meV. The $m=2$ distribution that then arises shows a series of bands with increasing energy in which the width of the distribution oscillates.  The $m=1$ distribution shows this oscillation too, but less obviously than for $m=2$.

To help understand these results, we consider the case when the maximum number of phonons that can be emitted is $M=1$.  This occurs when
$
  \frac{1}{2} \hbar \Omega + \hbar \omega_\text{LO}
  \le 
  E^{(0)}
%   -\epsilon_1
  < 
  \frac{1}{2} \hbar \Omega + 2\hbar \omega_\text{LO}
$.
In this case, \eq{EQ:EOM_Pall} can be solved exactly. With Gaussian initial conditions, the one-phonon distribution for subband $n$ reads
\beq
  A^{(1)}_n(x,t) 
  &=& 
  \frac{\Gamma_{n0}^{(0)} v^{(1)}_n}{2(v^{(0)}_0-v^{(1)}_n)}
  \left\{
     e^{-\Gamma_0^{(0)}t} f(x-v_0^{(0)} t)C(z_2)
   \right.
     \nonumber
     \\
   &&~~~~~~~~~~~~~~
   \left.
    -f(x-v_n^{(1)} t) C(z_1)
  \right\}
  \label{EQ:A1}
  ,
\eeq
with continued fraction ($z>0$) \cite{Jones1985}
\beq
  C(z) &=& 
  \exp(z^2)\mathrm{erfc}(z)
  \nonumber\\
  &=&
  \frac{2z}{\sqrt \pi}
  \cfrac{1}{2z^2+1-
    \cfrac{1\cdot2}{2z^2+5-
      \cfrac{3\cdot4}{2z^2+9-\cdots
      }}}
,
\eeq
and
\beq
  z_1 &=& 
  \frac{
    \sigma^2 \Gamma_0^{(0)}
    -
    (v^{(0)}_0- v_n^{(1)})(x- v^{(0)}_0t)
  }{
  \sqrt{2} \sigma (v^{(0)}_0- v^{(1)}_n)
  }
  ;
  \nonumber\\
  z_2 &=&
  \frac{
    \sigma^2 \Gamma_0^{(0)}
    -
    (v^{(0)}_0- v^{(1)}_n)(x- v^{(1)}_nt)
  }{
  \sqrt{2} \sigma (v^{(0)}_0- v_n^{(1)})
  }
  .
\eeq
Taking the limit $\Gamma_0^{(0)} t \gg 1$, and approximating $C(z) \approx 1/(\sqrt{\pi} z)$ we obtain
\beq
  A^{(1)}_n(x,t) \approx \frac{\Gamma_{n0}^{(0)}}{\Gamma_0^{(0)}} f(x- v_{n}^{(1)} t)
  .
\eeq
The width of this distribution for edge channel $n$ is 
\beq
  \Delta \tau^{(1)}_{n} 
  \approx \Delta \tau_\text{i} \frac{v_0^{(0)}}{v_n^{(1)}} 
  %\ge  \Delta \tau_\text{i}
  \label{EQ:partialwidths}
  .
\eeq
Thus, we expect the complete $m=1$ ATD to be approximately given by a sum of Gaussians with centre and widths determined by the velocity ratios $v_0^{(0)}/v_n^{(1)}$ and weighted by the branching ratio $\Gamma_{n0}^{(0)}/ \Gamma_0^{(0)}$.

\begin{figure}[t]
  \psfrag{dt}{$\Delta \tau^{(m)}/\Delta \tau_\text{i}$}
  \psfrag{cack}{$E^{(0)}$~[meV]}
  \psfrag{B=5}{$B=5$T}
  \psfrag{B=2}{$B=2$T}
  \psfrag{m = 1}{$m=1$}
  \psfrag{m = 2}{$m=2$}
  \begin{center}
     \includegraphics[width=\columnwidth,clip=true]{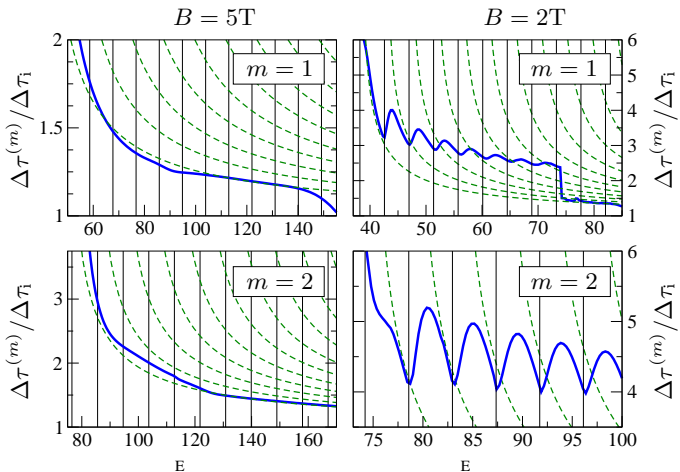}
  \end{center}
  \caption{  
    Widths $\Delta \tau^{(m)}$ of the arrival-time distributions for $m=1$ (top row) and $m=2$ (bottom) as a function of initial energy $E^{(0)}$ for two values of the magnetic field $B=5$T (left) and $B=2$T (right).
    The thick lines show the widths of the distributions;
    the green dashed lines show the partial widths $\Delta \tau^{(m)}_n$ predicted by \eq{EQ:partialwidths2};
    the black vertical lines show the (initial) energies for which $E^{(m)}$ is equal to a  subband-bottom energy, i.e. when $E^{(0)} = m \hbar\omega_\text{LO} + \hbar \Omega\rb{n+\frac{1}{2}} $.
    In all four panels, the leftmost partial width is that for $n=0$, with $n$ increasing in steps of one from left to right.
    At $B=5$T, the total width is determined almost exclusively by the properties of the outermost edge channel.  At $B=2$T, the width shows oscillations with increasing field arising from the contribution of a succession of different subbands. 
    \label{FIG:widths}
  }
\end{figure}

\fig{FIG:widths} shows the total widths $\Delta \tau^{(m)}$ of the distributions in Figs.~\ref{FIG:AxtB5} and \ref{FIG:AxtB2} in comparison with the partial widths 
\beq
  \Delta \tau^{(m)}_{n} 
  \approx \Delta \tau_\text{i} \frac{v_0^{(0)}}{v_n^{(m)}} 
  %\ge  \Delta \tau_\text{i}
  \label{EQ:partialwidths2}
  ,
\eeq
generalised from \eq{EQ:partialwidths}.
For $B=5$T, the situation is rather simple: across most of the energy range considered here, the full width decreases with $E$ and corresponds very closely to the $n=0$ partial width.  This is  indicative of the fact that, for these parameters, scattering is dominated by the $n=0 \to n'=0$ transition.  At higher energies, the width decreases below this level.  This effect has the same origin as the reduction in discussed in connection with \fig{FIG:P0}c.

The distribution widths for $B=2$T show oscillations  as a function of energy.  Analogously to the Shubnikov-de Haas oscillations, these occur as the relevant energy, here $E^{(m)}$, passes through the band-bottom energies. 
Consider first the $m=1$ case.  Below 
$
E^{(0)} = \frac{3}{2}\hbar \Omega + \hbar \omega_\text{LO}\approx42.5
$meV the only available state into which the electron can scatter is the outermost edge channel.  The width of distribution therefore follows $\Delta \tau_0^{(1)}$. 
At $E^{(0)}\approx42.5$meV, the band bottom of $n=1$ state passes through the energy $E^{(1)}$ and scattering into this state becomes possible.  Initially the rate of this process is small, but as energy increases, an increasing fraction of population is scattered into this state.   As the velocity of this $n=1$ state is significantly lower than that of $n=0$ state, the width of its partial distribution is greater.  Thus, as more population is added to this state, the width of distribution comes to be dominated by this wider contribution. Above a certain point the total width of the $m=1$ distribution follows roughly the $n=1$ partial width.  At 
$
  E^{(0)} = \frac{5}{2}\hbar \Omega + \hbar \omega_\text{LO}\approx47
$meV, the same thing occurs with the $n=2$ subband, which then determine the total width.  This pattern continues with increasing $E^{(0)}$.  At higher energies, however, the total distribution is actually a mixture of contributions from all the states with subbands below $E^{(1)}$. These different contributions combine to make the alignment of the oscillation minima drift away from the exact subband energies.

The $m=1$ width shows a sharp drop to the width of the $\Delta \tau_0^{(1)}$ level at 
$
  E^{(0)} = \frac{1}{2}\hbar \Omega + 2 \hbar \omega_\text{LO}\approx74
$meV.  This corresponds to the transition in the maximum phonon number from $M=1$ to $M=2$ when two-phonon processes become possible.  
In accordance with \fig{FIG:rates_m_ge_0}, the rates out of the $n>0$ states are greater than those from the $n=0$ states.  Thus, at these energies, any residual population in $m=1$ distribution will predominantly be in the $n=0$ state. At the onset of the $m=2$ distribution, its width starts below $\Delta\tau_0^{(2)}$.  This is a result of the admixture of components travelling with velocities from the source level (here $m=1$), just as \eq{EQ:A1} shows the $m=1$ distribution contains an admixture of slower $m=0$ velocity components.

Due to the passing of the electron energy through subband bottoms, the prediction of this effect for experiment can be only qualitative, rather than quantitative. Certainly near the bottom of the bands, the shape of the partial width curves will change from those shown here, since the dispersion relation will be modified.  However, when $E^{(m)}$ is near a subband bottom, the width is actually determined by the width of the next-lowest band, for which our electronic states should be a good approximation.  By the time the energy is increased enough to follow the partial width curve of a subband, the energy is already significantly higher than the band bottom. The width oscillations should therefore show a similar pattern in the open-parabola potential.

%%%%%%%%%%%%%%%%%%%%%%%%%%%%%%%%%%%%%%%%%%%%%%%%%%%%%%%%%%%%%%%%
%%%%%%%%%%%%%%%%%%%%%%%%%%%%%%%%%%%%%%%%%%%%%%%%%%%%%%%%%%%%%%%%

%%%%%%%%%%%%%%%%%%%%%%%%%%%%%%%%%%%%%%%%%%%%%%%%%%%%%%%%%%%%%%%%%%%%%%%%%
%%%%%%%%%%%%%%%%%%%%%      DISCUSSION     %%%%%%%%%%%%%%%%%%%%%%%%%%%%%%%
%%%%%%%%%%%%%%%%%%%%%%%%%%%%%%%%%%%%%%%%%%%%%%%%%%%%%%%%%%%%%%%%%%%%%%%%%
\section{Discussion \label{SEC:discussions}}

The advent of single-electron sources coupled with energy- and time-resolved detection opens new possibilities to study confined electron-phonon interactions in great detail \cite{Fletcher2013, Waldie2015}.
Here, we have calculated relaxation rates due to LO-phonon emission in quantum Hall edge channels and studied the consequences of these processes for the arrival-time distributions of electrons at a downstream detector.
The rates show an exponential suppression with increasing field strength as the electronic wave functions become more localised.  This, in turn, translates into a double-exponential suppression of the survival probability as the field decreases. 
These calculations explain the observations of phonon-scattering suppression at high magnetic fields in Refs.~\onlinecite{Fletcher2013, Waldie2015}.  
More quantitatively, \citer{Fletcher2013} reports an electron emission energy of 
$150$meV above the Fermi sea. Taking this to coincide with the energy above the potential bottom, our calculations suggest the magnetic-field at which the survival probability drops to one half is $B_{1/2} \approx 9$T for a detector at $x_\text{D}=3\mu$m.  This is consistent with the strong phonon scattering at $B=6$T and its almost complete absence at $B=12$T as observed in \citer{Fletcher2013}.

For scattering out of the higher, $n>0$ edge channels, our rates show an anomalously high value whenever the transition is vertical in the electron forward momentum.  In the experiments of Refs.~\onlinecite{Fletcher2013, Waldie2015}, it is believed that the electrons are emitted only into the $n=0$ level. To observe these anomalously high rates then, it is necessary to scatter electrons into these higher states post emission.  Phonon emission can not provide this scattering because, looking at \fig{FIG:rates_m_ge_0}, the crossing occurs at high field where scattering from the $n=0$ channel is both highly suppressed as well as effectively diagonal in subband index.  To observe these rates an additional scattering process, such as at a quantum point contact to populate inner edge channels, is required.

Concerning the ATDs themselves, we predict here an increase in the width of the phonon-replica distributions relative to that of the direct distribution.  Since this change in width depends on the change in $k$ value of the electron during phonon emission, at high energy, where the relative proportion of energy lost is small, the width increase will be correspondingly small.  However, at lower energy and field, this width increase can be significant.  In \fig{FIG:widths}, for example, the width of the $B=2$T distribution around their mid-point in energy range is a factor of three ($m=1$) or five ($m=2$) greater than the starting width.  Such increases should be visible with the time-domain resolution reported in \citer{Fletcher2013} and might prove a useful way to characterise the resolution of the detector.  
We have also seen that, at lower field, the distributions exhibit an oscillatory behaviour due to subband crossings.  The width of the distribution where these oscillations occur is significantly greater than the original emission width and should therefore be resolvable in current experiment.

There are a number of ways in which our calculations could be extended.  Firstly,   the description of states near the subband bottoms could be brought closer to those in experiment by considering the open-parabola potential.  Whilst this should give better agreement with experiment in specific parameters regions, some of the simplicity of the above approach will inevitably be lost.
More interesting will be to include further effects relating to the phonons.  In particular,  confinement of the optical phonons has been reported to be important in nanostructures in magnetic fields \cite{Telang1993}.  Moreover, it will be of interest to go beyond the Fermi-golden-rule approach used here and look for signatures of polaron physics in the single-electron arrival time distributions.

\begin{acknowledgments}
  A.~D. was supported by the Office of Naval Research sponsored by Dr. Paul Maki grant no. N00014-15-1-2193.
  H.~S.~S. was supported by Korea NRF (Grant No. 2013R1A2A2A01007327).
  M.~K. was supported by the UK Department for Business, Innovation and Skills, and NPL's Strategic Research Programme.
\end{acknowledgments}

%%%%%%%%%%%%%%%%%%%%%%%%%%%%%%%%%%%%%%%%%%%%%%%%%%%%%%%%%%%%%%%%%%%%%%%%%
%%%%%%%%%%%%%%%%%%%%%      APPENDIX     %%%%%%%%%%%%%%%%%%%%%%%%%%%%%%%
%%%%%%%%%%%%%%%%%%%%%%%%%%%%%%%%%%%%%%%%%%%%%%%%%%%%%%%%%%%%%%%%%%%%%%%%%
\appendix
\input{appendix}

%%%%%%%%%%%%%%%%%%%%%%%%%%%%%%%%%%%%%%%%%%%%%%%%%%%%%%%%%%%%%%%%%%%%%%%%%
%%%%%%%%%%%%%%%%%%%%%      REFERENCES     %%%%%%%%%%%%%%%%%%%%%%%%%%%%%%%
%%%%%%%%%%%%%%%%%%%%%%%%%%%%%%%%%%%%%%%%%%%%%%%%%%%%%%%%%%%%%%%%%%%%%%%%%

\input{phonon_rates_theory_v4.bbl}
\end{document}

%% file: appendix.tex
\section{Quantum hall wave functions \label{SEC:APQHstates}}

The eigenfunctions of the Hamiltonian of \eq{EQ:h} are
\beq
  \psi_{nk}(x,y) 
  = 
  \frac{1}{\sqrt{L_x l_\Omega}} e^{i k x}u_n\rb{\frac{y- y_G}{l_\Omega}}
  ,
\eeq
where normalisation in the $x$-direction is to the length of the conductor, $L_x$, and
where
\beq
  u_n(s) =  \frac{1}{\sqrt{2^n n!}}\frac{1}{\pi^{1/4}}e^{-s^2/2}H_n(s)
  ,
\eeq
are the standard harmonic-oscillator functions with $H_n(s)$ the $n$th Hermite polynomial.

To calculate the interaction with bulk phonons, we supplement this wave function with that in the growth direction, $z$. Assuming the confinement in this direction to be an infinite-square-well with boundaries at $z=0$ and $z=a$, the ground-state wave function in this direction is
\beq
  \phi_{1}(z) = 
  \begin{cases} 
     \sqrt{\frac{2}{a}} \sin \frac{\pi z}{a} &\mbox{for } 0\le z \le a; \\
     0 & \mbox{otherwise}.
  \end{cases}
\eeq
We assume that the energy of excitation in this $z$ direction is large enough that the electron is confined to this ground state.  The total wave function used in the calculation of matrix elements is then
\beq
  \Psi_{nk}(x,y,z) = \psi_{nk}(x,y) \phi_{1}(z)
  \label{EQ:fullwavefn}
  .
\eeq

\section{Calculation of electron-phonon rates\label{SEC:APep}}

Specified in terms of $\tilde{c}_\mathbf{k}$, the annihilation operator for plane-wave electrons with three-dimensional wave vector $\mathbf{k}$, the Fr\"{o}hlich Hamiltonian for polar optical phonons reads \cite{Mahan2000} 
% (NB: I think there is a typo in Mahan Eq.~(6.1.1) with the labels on the phonon operators; this is the form from page 44, which I believe is correct. Angela warns ``SI vs cgs units?'') 
\beq
  V_\text{ep}
  = 
  \sum_{\mathbf{k},\mathbf{q}} 
    M(\mathbf{q})\,
    \tilde{c}_{\mathbf{k}+\mathbf{q}}^\dag \tilde{c}_\mathbf{k}
    \rb{a_{-\mathbf{q}}^\dag  + a_\mathbf{q}}
  \label{EQ:FH1}
  ,
\eeq
with matrix element
\beq
  M(\mathbf{q}) 
  \equiv  
  \frac{M_0}{\sqrt{V}} 
    \frac{1}{|\mathbf{q}|}
    ;\quad
  M_0^2 = 4\pi \alpha \hbar \frac{\rb{\hbar \omega_\text{LO}}^{3/2}}{\rb{2m_e^*}^{1/2}}
  \label{EQ:M0alpha}
  .
\eeq
Translating into the basis of \eq{EQ:fullwavefn}, we obtain \eq{EQ:V} with the matrix elements \cite{Scher1966}
\beq
  \Lambda_{n'n}^{k'k}(\mathbf{q})
  &=&
  \sum_{\mathbf{p}} 
    M(\mathbf{q})
    \ew{\Psi_{n'k'}|\mathbf{p}+\mathbf{q}} 
    \ew{\mathbf{p}|\Psi_{nk}}
    ,
  \nonumber\\
  &=&
  M(\mathbf{q}) \,
  \delta_{q_x, k'-k}
  G^{(y)}_{n'k',n k}(q_y)
  G^{(z)}_{11}(q_z)
  ,
\eeq
with 
\beq
  G^{(y)}_{n'nk}(q_y) 
  &=&
%   \int dy e^{iq_y y} \chi_{n'k'_x}^*(y)\chi_{nk}(y),
  l_\Omega^{-1} e^{-iq_y y_G}  
  \int dy \, e^{iq_y y} 
    u_{n'}^*(l_\Omega^{-1}y)
    u_{n}^*(l_\Omega^{-1}y)
  ;
  \nonumber \\
   G^{(z)}_{11}(q_z) 
   &=& 
   \int dz \, e^{i q_z z} \phi^*_{1}(z)\phi_{1}(z)
   .
\eeq

Fermi's golden rule then gives the transition rate from state $n$ with energy $E$ to state with $n'$ as
\beq 
  \Gamma_{n'n}(E) 
%   &=& 
  =
  \frac{L_x}{\hbar}
  \int dk'
  \sum_\mathbf{q} 
  |\Lambda^{k'k}_{n'n} (\mathbf{q})|^2
%   \nonumber\\
%   &&
%   ~~~~~~~~~~~~~~~~
%   \times
  \delta\rb{E_{n'k'} - E + \hbar \omega_\text{LO}}
  .
  \nonumber
  \\
  \label{EQ:FGR}
\eeq
Taking the continuum limit for $\mathbf{q}$, three of the four integrals in this expression can be evaluated analytically \cite{Ridley2009}. This yields \eq{EQ:Ifull} with the remaining integral
\beq
  I_{n'n}(\delta_G)
  &=& 
  \int dQ
%   \times
  F^{(z)}_{11}\rb{\sqrt{Q^2 + \rb{\frac{a \Omega}{l_\Omega \omega_c}}^2\delta_G^2}}
  \nonumber
  \\
  &&
  ~~~~~~~
  \times
  F^{(y)}_{n'n}
  \rb{\sqrt{
    \frac{1}{2}\left[
     \frac{l_\Omega^2}{a^2} Q^2
     + \delta_G^2
    \right]
  }}
  \label{EQ:APInt}
  .
\eeq
Here we have defined
\beq
  F^{(z)}_{11}(A)\equiv \int dQ \frac{1}{A^2 + Q^2}|G^{(z)}_{11}(Q/a)|^2
  \label{EQ:Ldef}
  ,
\eeq
which evaluates as
\beq
  F_{11}^{(z)}(A)
  = 
  \frac{
    3 \pi A^5 + 20 \pi^3 A^3 - 32 \pi^5(1-e^{-A})
  }
  {
    A^3(A^2 + 4 \pi^2)^2
  }
  ,
\eeq
and we have written
\beq
  |G^{(y)}_{n'n k}(q_y)|^2
  =
  F^{(y)}_{n'n}
  \rb{\sqrt{
    \textstyle{\frac{1}{2}}\left[
     l_\Omega^2 q_y^2 
      + \delta_G^2
    \right]
  }}
  ,
\eeq
with
\beq
  F^{(y)}_{n'n}(Q) = \frac{n_<!}{n_>!}e^{-Q^2} Q^{2|n'-n|}\left[L_{n_<}^{|n'-n|}\rb{Q^2}\right]^2
  \nonumber
  .
  \\
\eeq
In this latter, $L_n^\alpha(x)$ is an associated Laguerre polynomial and $n_<=\mathrm{min}(n_2,n_1)$ and $n_>=\mathrm{max}(n_2,n_1)$. 
In writing \eq{EQ:APInt}, we have neglected processes which change the sign of $k$.  
This is consistent with only one side of the potential being relevant to the experimental potential.  Processes that change the sign of $\mathbf{k}$ will anyway be severely suppressed when $y_G \gg l_\Omega$.

The approximation to  $I_{n'0}$ given in \eq{EQ:Iapp} can be obtained by noting that the exponential factor in $F^{(y)}$ dominates the integrand.  Setting $Q$ everywhere but in this exponent, we obtain
\beq
  I_{n'n}(\delta_G)
  &\approx& 
  \int dQ   e^{- \frac{1}{2}\rb{\frac{l_\Omega Q}{a}}^2 }
%   \times
  F^{(z)}_{11}\rb{\frac{a\Omega}{l_\Omega \omega_c}\delta_G}
  F^{(y)}_{n'n}
  \rb{\frac{1}{\sqrt{2}} \delta_G}
  \nonumber\\
  &=&
  \frac{\sqrt{2\pi}a}{l_\Omega}
  F^{(z)}_{11}\rb{\frac{a\Omega}{l_\Omega \omega_c}\delta_G}
  F^{(y)}_{n'n}
  \label{EQ:IQapp1}
  .
\eeq
% We will call this ``approximation 1''.  
Further, if the initial energy of the electron, $E$, is the dominant energy scale, we can approximate
\beq
  \delta_G \approx 
  \frac{\omega_c^2 \omega_\text{LO}}{\omega_y^2 \Omega}\frac{l_\Omega}{y_G}
  .
\eeq
This gives $\delta_G \sim E^{-1/2}$ and we thus expand the non-exponential contribution for small $\delta_G$ to obtain \eq{EQ:Iapp}.

\section{Quantum dynamics \label{SEC:APQM}}

We can assess the importance of dispersion by considering the time evolution of a Gaussian wave packet \cite{Pauli2000,Andrews2008}.
We assume that initially the electron is confined to the lowest Landau level ($n=0$) with momentum distributed normally. The wave function then reads
\beq
  \ket{\Psi(0)} = \sum_{k} N_\alpha e^{-\alpha(k-k_0)^2} \ket{\psi_{0k}}
  ,
\eeq
with $k_0$ the central wavenumber and $\alpha = \sigma^2$ a width parameter with $k_0 \gg (4\alpha)^{-1/2}$, such that only positive values of $k$ are relevant. $N_\alpha$ is a normalisation constant.
The marginal probability distribution of this state in the $x$-direction is a Gaussian with width
\beq
  \sqrt{\nu} = \sqrt{
    \alpha + \frac{1}{2}\rb{\frac{\omega_c}{\Omega} l_\Omega}^2 
  }
  .
\eeq
The increase over $\sqrt{\alpha}=\sigma$ stems from the dependence on wave number of the guiding centre of the transverse wave function.

The time evolution of this wave packet including phonon-induced relaxation can be found by solving the 
time-dependent Schr\"{o}dinger equation with eigen-energies that have imaginary parts 
to account for the population decay. The wave function at later time $t$ is then
\beq
  \ket{\Psi(t)} = \sum_{k} N_\alpha 
  e^{-\alpha(k-k_0)^2 - i \hbar^{-1} \widetilde{E}_{0k} t} 
  \ket{0, k}
  \label{EQ:APwfnt}
  ,
\eeq
with
$
  \widetilde{E}_{0k} = E_{0k} - i\frac{\hbar}{2}\Gamma_0^k
$ where, for the purposes of this appendix, we have labelled the rate with the initial wave number: $\Gamma_0^k = \Gamma_0(E_{0k})$.
At high energy and field, the rate is a slowly-varying function of $k$. We therefore approximate 
$
  \Gamma_0^k \approx \Gamma_0^{k_0}
  + \partial_k\Gamma_0^{k_0}\left[k-k_0\right] 
  + \frac{1}{2} \partial^2_k\Gamma_0^{k_0}\left[k-k_0\right]^2
$.

The simplest way to define an arrival time distribution in quantum mechanics is in relation to the current density \cite{Dumont1993,Delgado1999}
\beq
  A(x_\text{D},t)
  \equiv
  \int dy \, \mathbf{J}(x_\text{D},y,t)\cdot \mathbf{\hat{i}}
  ,
\eeq
where $\mathbf{J}(x,y,t)$ here is given by
\beq
  \mathbf{J}(x,y,t) 
  &=& 
  \frac{1}{m_e^*}
  \text{Re}
  \left\{  
    \Psi^*(x,y,t) 
      \rb{
        \mathbf{p} + \frac{e}{c} \mathbf{A}    
      }
    \Psi(x,y,t)
  \right\}.
  \nonumber\\
  \label{EQ:Jdefn}
%   .
\eeq
This definition can lead to problems of interpretation \cite{Allcock1969,Muga2000} but under the conditions studied here, no such issues arise.
With the wave function of \eq{EQ:APwfnt}, we obtain
\beq
   A(x_\text{D},t)
   &=&
  \frac{1}{ \sqrt{2\pi \widetilde{\nu}}}
   D^{-3/2}
   \tilde{v}_g
   \left[ 
     1 + \rb{D-1}\frac{x}{\tilde{v}_g t}
   \right]
   \nonumber\\
   &&
   \times
   \exp
   \left[
     -\frac{1}{2 \tilde{\nu}D}\rb{x_\text{D}-\tilde{v}_g t}^2
   \right]
   \nonumber\\
   &&
   \times
   \sqrt{\frac{\alpha}{\tilde{\alpha}}}
   \exp
   \left[
     - \widetilde{\Gamma} t 
   \right]
   ,
\eeq
with
\beq
  \widetilde{\alpha} &=& \alpha + \frac{1}{4}\rb{\partial_k^2 \Gamma_0^{k_0}}t
  ;\quad\quad
  \widetilde{\nu} =  \widetilde{\alpha} + \frac{1}{2}\rb{\frac{\omega_c}{\Omega}}^2 l_\Omega^2
  ;
  \nonumber\\
  D &=& 1+ \frac{\hbar^2 t^2}{4 {m_e^*}^2\tilde{\alpha}\tilde{\nu}}
  ;\quad\quad
  \tilde{v}_g = v_g - \rb{\partial_k \Gamma_0^{k_0}}
    \frac{ \hbar t}{4m_e^* \tilde{\alpha}}
  ;
  \nonumber\\
  \widetilde{\Gamma}&=& \Gamma_0^{k_0} + \rb{\partial_k \Gamma_0^{k_0}}^2\frac{ t}{8 \tilde{\alpha}}
  .
\eeq
This can be understood as the ATD of a damped, travelling Gaussian wave packet moving with velocity, width and damping rate that are all functions of time.

% \subsection{relevance to exp.}
We can assess the importance of the time-dependence of these parameters for experiment. By assuming $t \sim x_\text{D}/v_g$, a width parameter $\alpha \approx (1\mu\text{m})^2$ and detector position $x_\text{D} = 10\mu$m, we numerically obtain a value of
$\tilde{\alpha}/\alpha - 1 \approx 10^{-5}$ for $B=5$T and $E=100$meV, which is a typical value in the relevant parameter range.  A similar story can be told for the corrections involving the derivatives of the rates --- these terms yield relative corrections of the order of $10^{-5}$ or less. We are thus safe in approximating $\tilde{\alpha} \approx \alpha$, $\tilde{\nu} \approx \nu$, $\tilde{\Gamma} \approx \Gamma_0^{k_0}$ and $\tilde{v}_g \approx v_g$.

Concerning the width adjustment, we have
\beq
  \frac{\tilde{\nu}}{\tilde{\alpha}} - 1 
  \approx 
  \frac{\nu}{\alpha} - 1 
%   = 
%   \frac{1}{2} \rb{\frac{\omega_c}{\Omega}}^2 \frac{l_\Omega^2}{\alpha}
%   =
%   \frac{1}{2}  \rb{1+\frac{\omega_y^2}{\omega_c^2}}^{-3/2} \frac{l_c^2}{\alpha}
  \sim  \rb{\frac{l_c}{\sigma}}^2
  ,
\eeq
with $l_c=\sqrt{\hbar/m_e^* \omega_c}$ the magnetic confinement length. At $5$T, $l_c \sim 10$nm, so this correction is also small.  Finally, we have the diffusion parameter
\beq
  D - 1 
%   \approx
%   \frac{\hbar^2 t^2}{4 {m_e^*}^2\alpha^2}  
  \approx
  \frac{\hbar^2 x_\text{D}^2}{4 {m_e^*}^2v_g^2\alpha^2} 
  \approx 10^{-5}
  ,
\eeq
and the dispersion of the initial wave packet is also negligible. 

Neglecting these small terms then, the arrival time distribution becomes that of a dispersionless Gaussian wave packet moving at velocity $v_g$ and damped at a rate $\Gamma_0$:
\beq
   A(x_\text{D},t)
   &=&
   \frac{1}{\sqrt{2\pi \alpha}}
  v_g  
  e^{-\Gamma_0^{k_0} t}
   \exp
   \left[
     - \frac{1}{2\alpha} 
     \rb{x-v_g t}^2
   \right]
   \nonumber
   .
\eeq
This agrees with the expression found in section \ref{SEC:survprob} and the lack of any significant corrections to this simple Gaussian evolution justifies the use of the semi-classical approach in the main text.